\documentclass{icrc29}
\usepackage{graphicx,amssymb,amsmath,times}
\begin{document}
\title[Atmospheric muon background in the ANTARES detector]{Atmospheric muon background in the ANTARES detector}
\author[S. Cecchini et al.] {S. Cecchini $^a$ $^b$, E. Korolkova  $^c$, A. Margiotta $^a$, L. Thompson $^c$, \\
\normalfont{on behalf of the ANTARES Collaboration}\\
        (a) Dipartimento di Fisica dell'Universita' di Bologna e Sezione INFN di Bologna, viale C. Berti Pichat, 6/2,\\ 40127 Bologna, Italy \\ 
        (b) Also  IASF/CNR, Bologna, Italy \\
        (c) Department of Physics and Astronomy, University of Sheffield, Hicks Building, Hounsfield Road, Sheffield S3 7RH, \\ United Kingdom}
\presenter{Presenter: A. Margiotta (annarita.margiotta@bo.infn.it), ita-margiottaA-abs1-og25-poster
 }
\maketitle
\begin{abstract}
An evaluation of the background due to atmospheric muons in the ANTARES high energy  neutrino telescope is presented. Two different codes for atmospheric shower simulation have been used. Results from comparisons 
between these codes at sea level and detector level are presented.
The first results on the capability of ANTARES to reject this class of background are given.
\end{abstract}
\section{Introduction}
The detection of neutrinos from galactic/extragalactic sources has enormous potential for the understanding of astrophysical processes and for an improved exploration of the far Universe.
Neutrino astronomy is assuming an important role in experimental physics research. The principle of detection in an underwater neutrino telescope is based on the emission of Cherenkov light by muons emerging from neutrino interactions inside and around the detector.\\
The ANTARES Collaboration \cite{antares} is  building an undersea neutrino telescope off the French coast at a depth of 2500 m. The telescope will consist of 900 photomultipliers (PMTs) arranged in 12 strings. The arrival times of Cherenkov photons on the PMTs will be registered with an accuracy of about 1 ns, which will result in an angular resolution of a few tenths of a degree for the reconstruction of a high energy muon track. 
In principle, selecting upward going muons would discriminate the muon tracks induced by neutrinos (both atmospheric and astrophysical). In practice, however,  hits due to Cherenkov light emitted by  the downward going atmospheric muons can be erroneously reconstructed as upgoing tracks.\\
A complete Monte Carlo simulation has been performed to study the response of the detector to atmospheric muons and to estimate the possible contamination from misreconstructed tracks, starting from the production of cosmic ray induced atmospheric showers, propagating the surviving muons through water to the detector, simulating the Cherenkov light production and the hits on the PMTs and, finally, reconstructing muon tracks.\\
Atmospheric shower production and propagation through the atmosphere has been performed using two different codes and the results have been compared.
\section{Atmospheric muon simulation.}
The first step of the simulation is the production of the atmospheric showers induced by primary cosmic rays. Two different programs have been used: HEMAS \cite{hemas} and CORSIKA v.6014 \cite{corsika}.\\
The HEMAS code was developed for the simulation of the atmospheric muon flux in the MACRO \cite{macro}  underground detector, at the Laboratori Nazionali del Gran Sasso. It takes into account all the main physical processes occurring in the atmosphere: it computes the first interaction point of primary cosmic rays on the basis of the input cross sections, propagates the electromagnetic and hadronic components of the showers, takes into account the deflection of charged particles due to the geomagnetic field and the Earth's curvature, allowing the calculations at large zenith angle. Each hadronic interaction in the atmosphere is handled with the hadronic interaction code DPMJET v.II \cite{dpmjet}. Some approximations in the particle transport along the atmosphere restrict the application of the code to particle energies greater than 500 GeV.\\
The CORSIKA package is the most widely used tool to model extensive air showers (EAS). It performs a detailed simulation of the evolution in the atmosphere of an EAS  initiated by various cosmic ray particles up to energies of some $10^{20}$ eV. Single particles are tracked, accounting for energy losses, deflection due to multiple scattering and the Earth's magnetic field, decays of unstable particles and electromagnetic and hadronic interactions. The CORSIKA code provides the option to choose one of several hadronic interaction models. In this work the QGSJET \cite{qgsjet} model has been chosen to treat high energy hadronic interactions.\\
To represent the primary energy spectrum, a simplified version of the  composition model described in \cite{horandel} was used. All the elements heavier than iron have been neglected and the remaining ones have been grouped in five mass groups: \textit{i)} protons; \textit{ii)} helium; \textit{iii)} CNO; \textit{iv)} Mg; \textit{v)} Fe and heavy nuclei.
For each nucleus, the production has been subdivided into several energy intervals ranging from 1 to $2 \times 10^6$ TeV and 2 zenith intervals: \textit{i)}	 $0^o < \vartheta < 60^o$; \textit{ii)} $60^o < \vartheta < 85^o$. The total number of simulated showers is about $7 \times 10^8$ for HEMAS  and $2 \times 10^8$ for CORSIKA. 
The corresponding detector livetime varies depending on the energy and zenith intervals chosen, going from some hours in the lower energy range to  years in the higher.

\begin{figure}[!h]
\hspace{-.4cm}
\includegraphics*[width=0.5\textwidth,angle=0,clip]{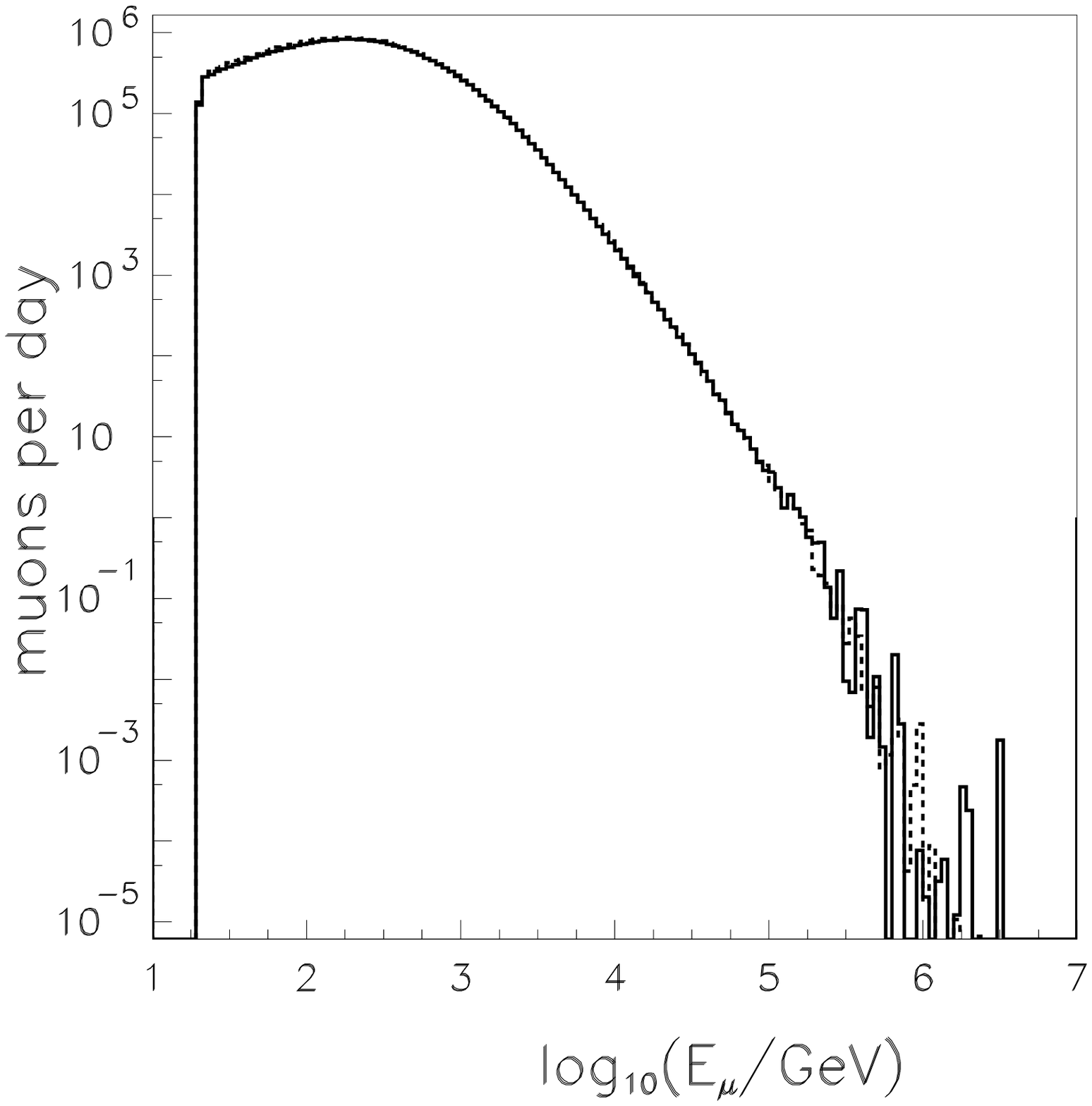}
\hspace{0.6cm}
\includegraphics*[width=0.5\textwidth,angle=0,clip]{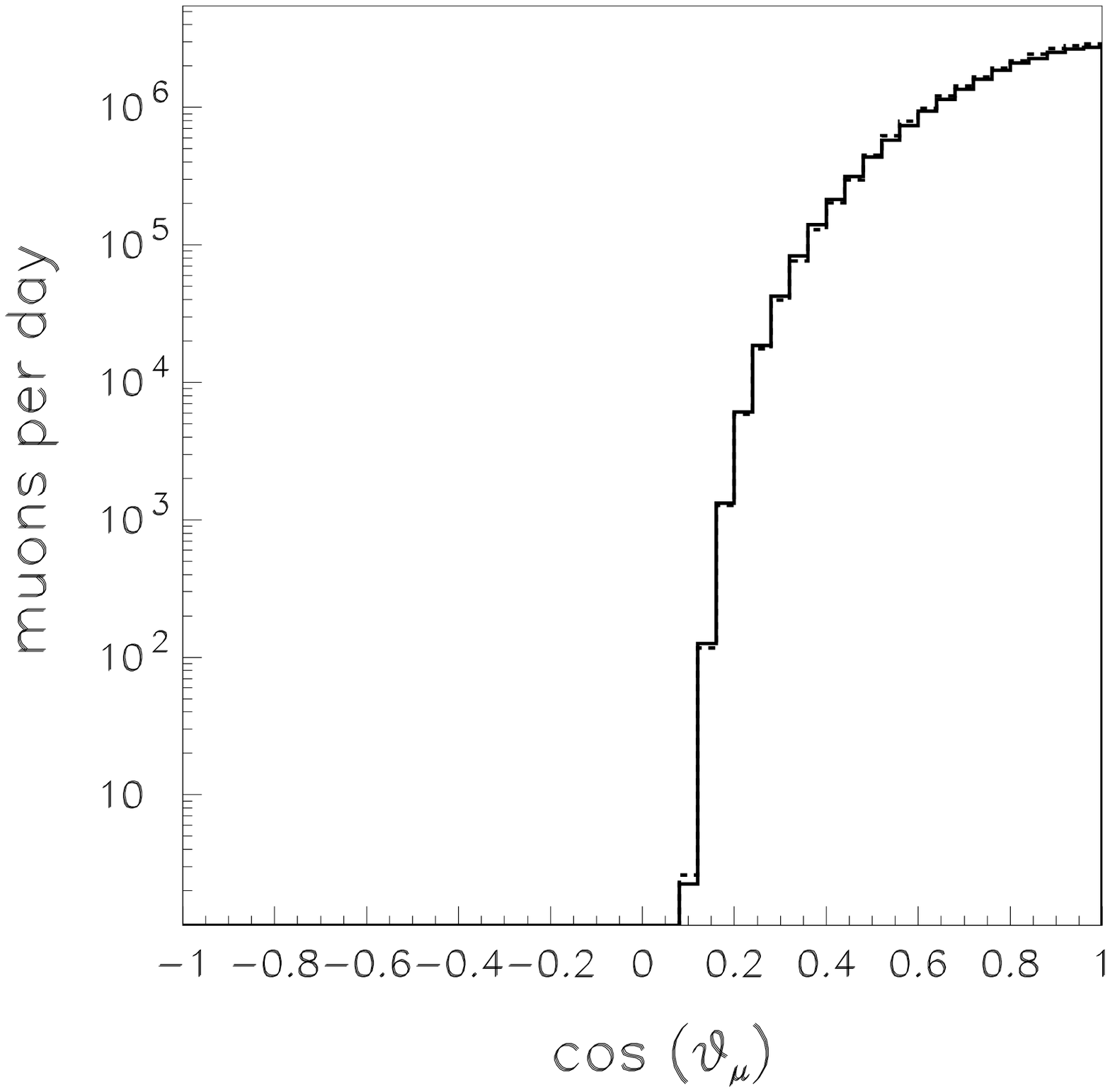}
\vspace{-1.cm}
\caption{Atmospheric muon flux arriving at the detector surface. Left: Muon energy spectrum. Right: muon angular distribution. The continuous line corresponds to the HEMAS simulation, the dotted line  to the CORSIKA simulation }
\label{fig:emu}
\end{figure}

Muon propagation through water is performed with  the MUSIC (MUon Simulation Code) program \cite{music}.
MUSIC  is a 3D muon propagation code that takes into account energy losses due to bremsstrahlung, pair production, inelastic scattering and ionization. It simulates also the angular and lateral deflections due to multiple scattering.
In order to sample the showers on the detector surface, a cylindrical volume, surrounding the sensitive volume of the detector, is defined around the PMT  array, the \textit{can}. Its dimensions are obtained adding $2.5 \times \lambda_{abs}$ to the maximum distance of PMTs from the centre of gravity of the PMT array, where $\lambda_{abs}\! = \: 55 \: m $ is the water absorption length at the ANTARES site.
The dimensions of the \textit{can} used in this work are : radius $\sim 240 $ m; height $\sim 590 $ m.\\
\noindent For each shower a random point is generated on the surface of a cylinder that is larger than the  \textit{can}, in order to take into account the lateral extension of the shower. Starting from this impact point, all the possible intersection points of each muon with  the  \textit{can} surface are evaluated. Each particle is then propagated along the path between the sea level and the detector using the MUSIC code, requiring a minimum residual energy of 20 GeV at the surface of the \textit{can}. \\
Figure 1 shows  the energy spectrum and the angular distribution, normalised to one day of live time, of the muons reaching the total \textit{can} surface. The continuous line, corresponding to the HEMAS simulation, and  the dotted line, relative to the CORSIKA simulation, are almost perfectly coincident. The total number of muons is about $2.7 \times 10^7$ with a difference smaller than $4\%$ between the two productions.\\
The third step of the simulation concerns the detector response and  is performed using a Geant3 \cite{geant3} based Monte Carlo code. An optical background rate of 60 kHz per PMT has been added. This is the rate expected from the presence of $^{40}K$ in sea water and has been measured in situ \cite{opt_bg}.\\
\section{Track reconstruction.}
Finally, the hits registered on the PMTs have been processed in order to reconstruct the track direction. Several strategies have been developed in ANTARES for track reconstruction. 
\begin{figure}[!h]
\hspace{-0.4cm}
\includegraphics*[width=0.5\textwidth,angle=0,clip]{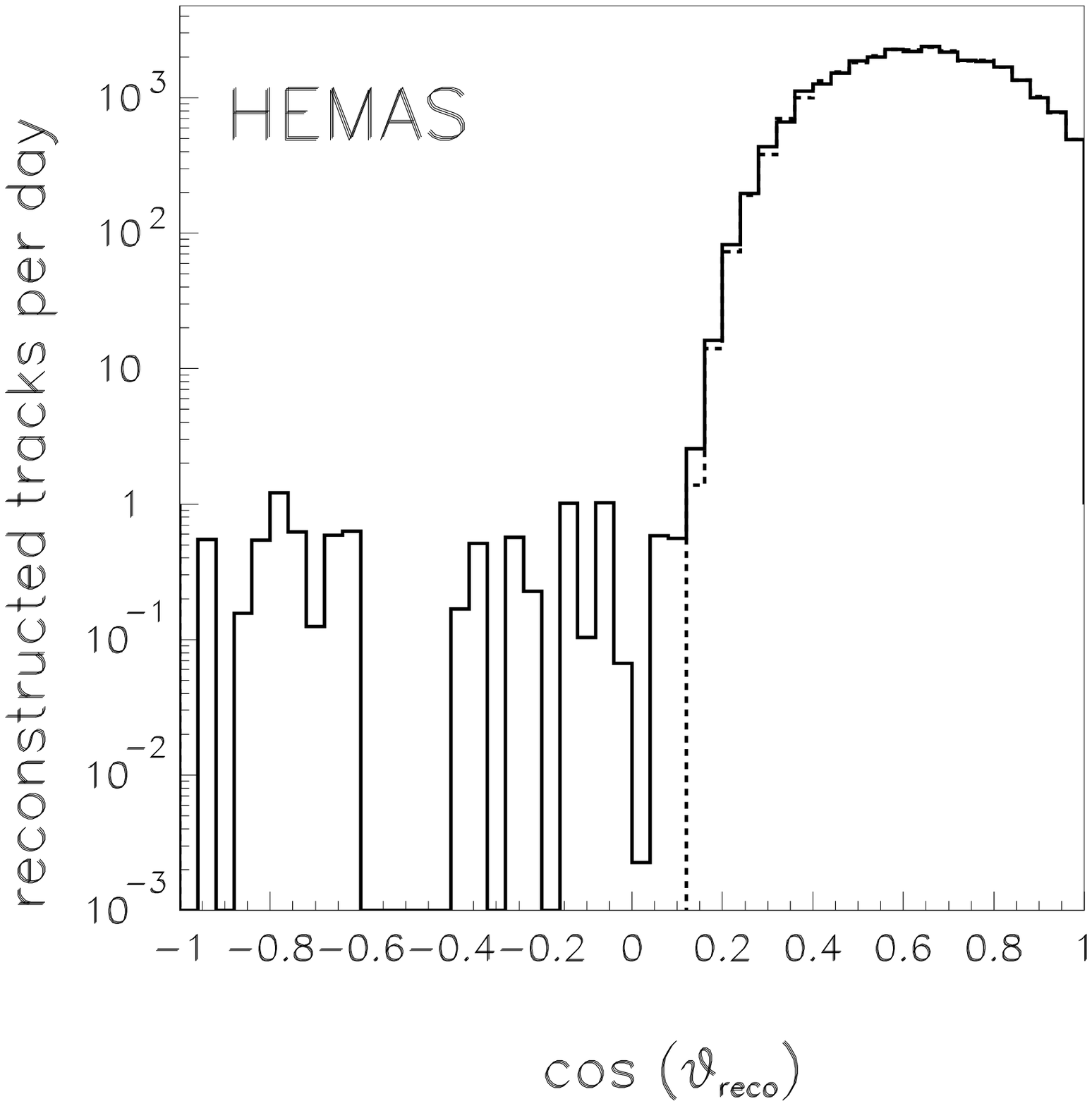}
\hspace{0.6cm}
\includegraphics*[width=0.5\textwidth,angle=0,clip]{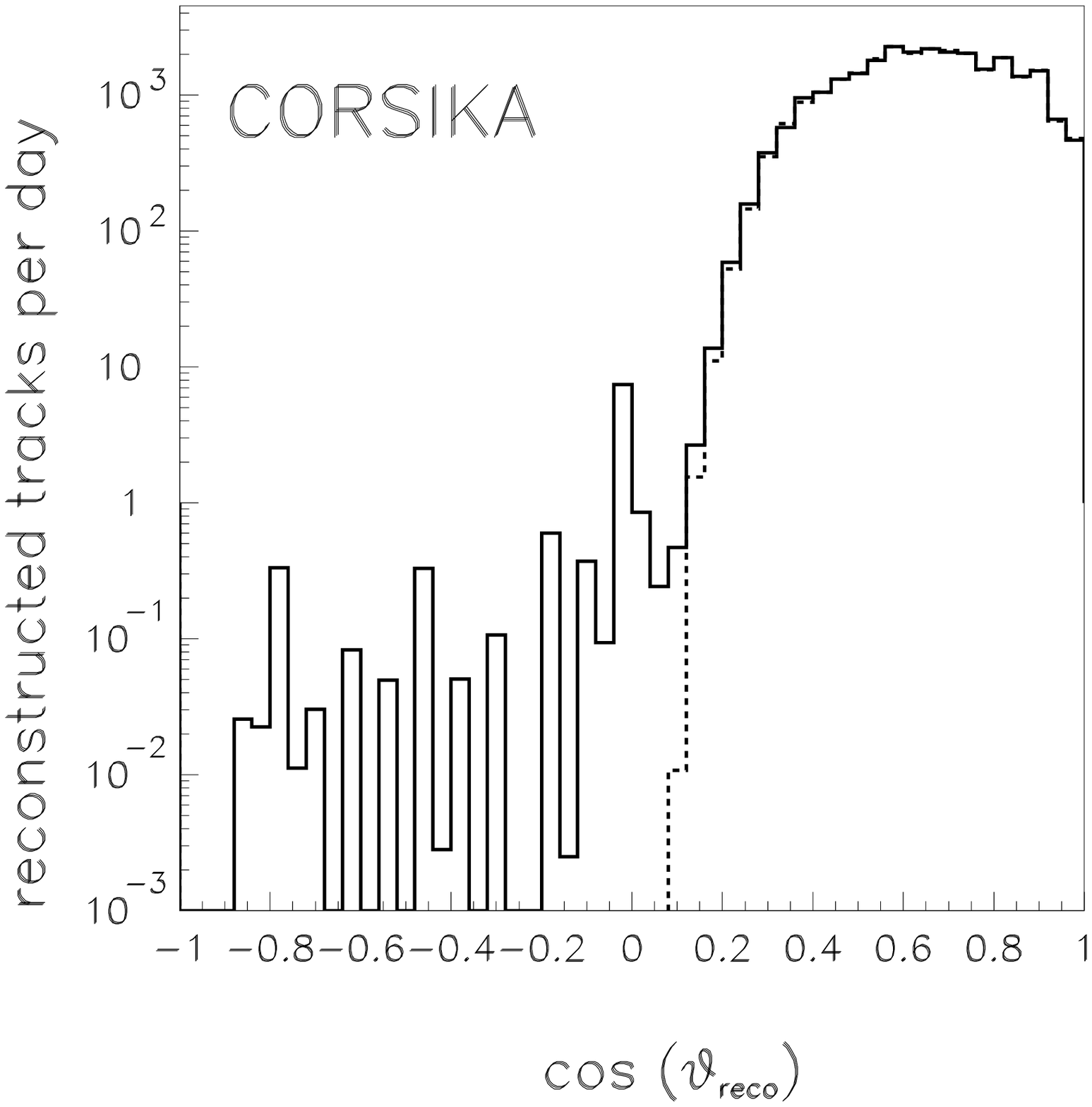}
\vspace{-.8cm}
\caption{Angular distribution of reconstructed muon tracks with the \cite{aart_th} strategy for: HEMAS sample (left);  CORSIKA sample (right). The continuous lines refer to the reconstructed angles, the dotted lines to the original direction of events.}
\label{fig:reco}
\end{figure}  In this work two approaches have been considered, both of them  based on the calculation of a likelihood function, starting from the recorded information of the arrival time and of the amplitude of hits on each PMT. Its maximum is used to obtain an estimate of the muon track parameters. \\
The first strategy  is described in detail in \cite{aart_th}. The results obtained when applied to the muon atmospheric samples from the HEMAS and CORSIKA simulations are shown in figure 2. The continuous lines refer to the reconstructed angles, the dotted lines to the original direction of events.
About $2.6 \times 10^4$ downward going muon tracks per day are reconstructed. Also some fake upward going events are reconstructed: $8 \pm 2$ per day for HEMAS  and $10 \pm 2$ for CORSIKA. 
About $40\%$ of them  correspond to muons arriving in a bundle.\\
The other strategy  used for reconstruction is described in \cite{carmona}. It is optimized to reconstruct upward going tracks and its efficiency in reconstructing downward going atmospheric muons is  poor with respect to the first strategy.  On the other hand, it has a better capability to reject misreconstructed upward going muons (no  upward going tracks have been reconstructed, when applied both to the  HEMAS and the CORSIKA samples).  

\section{Conclusions}
A full simulation of the atmospheric downward-going muon flux has been performed using two atmospheric shower simulation programs, HEMAS and CORSIKA. No significant differences have been found at the detector level (see fig. 1) between the two samples. \\
An evaluation of the contamination due to the atmospheric downward-going muons erroneously reconstructed as upward going is presented, using two reconstruction strategies available in the ANTARES Collaboration. 
One of them \cite{aart_th} shows a better efficiency in downward track reconstruction, while the other one \cite{carmona} rejects all upward going reconstructed tracks. The choice of the most appropriate strategy for reconstruction strongly depends on the characteristics of the analysis to be  performed (diffuse neutrinos,  pointlike neutrino sources,  atmospheric muons etc.). An excellent agreement between the two samples is obtained also after the reconstruction step.\\
The results are very encouraging. Work is in progress to improve their statistical significance, increasing the number of simulated events.

\newpage
\end{document}